\documentclass[aps,prl,twocolumn,superscriptaddress,nofootinbib,
nobalancelastpage,nobibnotes]{revtex4}
\usepackage{epsfig}
\begin{document}

\title
{Neutrino cloud instabilities just above the neutrino sphere of a supernova}

\author{R. F. Sawyer}
\affiliation{Department of Physics, University of California at
Santa Barbara, Santa Barbara, California 93106}

\begin{abstract}
 Most treatments of neutrino flavor-evolution, above a surface of the last scattering, take identical angular distributions
 on this surface for the different initial (unmixed) flavors, and for particles and antiparticles.  Differences
in these distributions must be present, as a result of the species-dependent scattering cross-sections lower in the star, 
These lead to a new set of non-linear equations, unstable even at the initial surface with respect to perturbations that break all-over spherical-symmetry. There could be important consequences for explosion dynamics as well as for the neutrino pulse in the outer regions.
\end{abstract}
\maketitle 

The study of collective neutrino interactions in the outskirts of the supernova explosion has generated an explosion of publications
 \cite{raf1}-\cite{duan2}. These collective effects come about as a consequence of the standard model neutrino-neutrino coupling, and change the correlations between neutrino flavor and neutrino energy. In doing so they
may profoundly affect the neutron-proton ratio, and R-process rates in layers far above that of the last scattering.

In these studies what is added to the observationally well established single $\nu$ oscillation and matter effects can be described as totally forward, standard model, interactions among the $\nu$, {\it viz.},  $\nu (\vec q)+ \nu(\vec k) \rightarrow \nu (\vec q)+ \nu(\vec k)$. Nothing except $\nu$-flavor enters into the process.
In principle, therefore, one can follow outward each neutrino, continuously updating the flavor density-matrix of each to take into account the interactions with all of the others.  Consideration of this process, as codified in a Liouville-type equation \cite{raf1}, has led to a number of large-scale numerical calculations, e.g. refs. \cite{fuller}-\cite{fuller9}, in which the underlying assumptions have been named the
``$\nu$-bulb" model.

These calculations begin at the surface of a neutrino-sphere, a loosely-defined surface beyond which scattering is considered ignorable, and they assume spherically symmetrical neutrino flows from below. On this $\nu$-surface  the $\nu$ angular distributions are taken as uniform, with no variation in angle over the outward directions, but with no inward bound $\nu$'s. The initial states for these neutrinos are taken as ``flavor-diagonal", that is with no mixing of $\nu_e, \nu_x$ or of $\bar \nu_e, \bar \nu_x$. Here $\nu_x$ is that combination $\nu_\mu$ and $\nu_\tau$ that is best suited to two-neutrino simulations.

There have been a number of recent works \cite{rfs} -\cite{duan5} devoted to the possibility of ``azimuthal" or ``multi-angle" instabilities in the 
above picture; that is, that some small perturbation of the $\nu$ flavor density matrix that breaks the spherical symmetry of system can grow exponentially beyond some radius of onset. In these examples the instabilities would occur at distances of 100 km. or more from the 
``$\nu$-surface.", a sphere of last scattering that is at a radius of about 15 km.. By contrast we find instability at or very near this surface.
There is a reason for this difference: In some imprecise way it is already known that the onset of a``multi-angle" instability, stemming from the $\nu-\nu$ interaction, requires a state in which angular distributions are flavor dependent \cite{rfs}, \cite{raf6}, and in the references cited above these develop through joint action of the $\nu-\nu$ interaction and neutrino oscillation terms. The latter are greatly suppressed by the interaction with electrons until the $\nu$'s are out to the 100 km. region. In contrast we begin at the $\nu$-surface with a physically based flavor dependence in the angular distribution that makes the system unstable to the tiniest seed from oscillation effects.

Some features of our approach are:\newline
1. We follow custom by taking distributions on initial surfaces to be angle-independent in the outgoing hemisphere. 
But we take the radii of these surfaces to depend on ``flavor", which in our terminology will distinguish $\{\nu_e,\bar \nu_e, \nu_x,\bar \nu_x \}$.
This is motivated by the fact that
the average last-scattering radius $r_{\nu_e}$, for $\nu_e$, is considerably larger than that for $\bar \nu_e$ which, in turn, should be larger than that for $\nu_x,\bar \nu_x$. 

2. Then we begin our real calculation (still with flavor-diagonal distributions) at an initial distance $r_{\nu_e}$. The other three favors on this surface have narrower angular cones than does $\nu_e$, just from the effects of free streaming from their lower surfaces of origin.

3. The distributions on an initial surface that are produced from the above are perfectly spherically symmetric in the sense that $\nu$ 
angular distributions, $f(\theta)$, where $\theta$ is the angle with respect to the local normal, are the same at all points of the sphere. Testing the non-linear evolution equations for stability against non-spherically symmetric perturbations in the flavor density matrices, we find instability at the initial surface at $r_0$ (with $r_0 >r_{\nu_e})$ for an interesting range of parameter values. 

 4. The system described above still has perfect spherical symmetry in its flavor distribution and flow. The perturbations that grow, at a rate of order $G_F n_\nu$ as it turns out, are characterized by having a combination of tiny flavor mixing and spherical asymmetry.  Seeds of such irregularities can come about in many ways. We mention one that does not even require asymmetry in the $\nu$ number flow: Neutrino oscillation terms will give
 tiny flavor mixings during the propagation between last scattering and the starting surface of the calculation. Irregularities in electron densities in this region will lead to angular irregularities in the starting flavor-density matrix, insignificant in stable cases but crucial in unstable cases.

We begin from the Liouville equation \cite{raf3},\cite{cardall} for time independent flow,
\begin{eqnarray}
& {\bf \hat n\cdot \nabla} \vec  \sigma_{\alpha} ({\bf r},{\bf \hat n})  = 2^{1/2}(2 \pi)^{-3}G_F   \vec \sigma_{\alpha} ({\bf r},{\bf \hat n})\,\times
\nonumber\\
& \sum_{\beta=1}^4 \zeta_\beta \int d\Omega_{\bf \hat n'}   \vec  \sigma_{\beta}  ({\bf r},{\bf \hat n'})
(1-{\bf \hat n} \cdot {\bf \hat n'} ) +\{ \rm O\&M\}\,,
\label{master0}
\end{eqnarray}
where the vectors indicated by arrows and the vector product are in $\nu$-flavor space and ${\bf \hat n}$ is neutrino velocity in real space. $\{ \rm O\&M\}$ stands for the standard oscillation and matter terms, linear functions of the components of $ \vec \sigma_{\alpha}  $, which we drop from the development below, since they will enter only in the preparation of the initial state for the calculation. 
The index $\alpha$ keeps track of distinguishing the origin of a neutrino as a pure flavor state
in one of the groups $\{\nu_e,\bar\nu_e,\nu_x, \bar \nu_x\}$. The set  $\zeta_\alpha=\{1,-1,1,-1\}$ puts in minus signs for anti-$\nu$'s inside the sum. In (\ref{master0}), and everything that follows, the  energy spectrum is irrelevant; it is number densities that enter. 

In the case of a spherically symmetric system we have $\vec\sigma_\alpha({\bf r, \hat n})= \vec\sigma_\alpha(r,{\bf \hat r \cdot \hat n})$,
whence,
\begin{eqnarray}
&[{\bf \hat n} \cdot   \nabla \, ] \vec \sigma_\alpha  ( r, { {\bf \hat r \cdot \hat n}}  )=
\nonumber\\
&{\bf \hat n} \cdot \Bigr [ {\bf \hat r}\, \partial_1 \vec \sigma_\alpha \Bigr( r,{\bf \hat r\cdot  \hat n} \Bigr )+
 {1\over r}({\bf \hat n}-{\bf \hat r[ \hat r \cdot \hat n}]\,)~\partial_2 \, \vec \sigma_\alpha \Bigr( r,{\bf \hat r \cdot \hat n}\Bigr )\Bigr ]
\nonumber\\
&=\Bigr [
 \cos\theta{\partial \over \partial r}+{\sin^2 \theta\over r}{\partial \over \partial \cos \theta}\Bigr ]
 \vec \sigma_\alpha (r,\cos \theta)\equiv \mathcal{O} \vec \sigma_{\alpha}(r,\cos \theta) ,
  \nonumber\\
  \,
 \label{angles}
   \end{eqnarray}
where $\partial_1$ and $\partial_2$ are the derivatives with respect to the first and second arguments of $ \vec \sigma_\alpha$,
 the angle $\theta$ is between ${\bf r}$ and $\hat n$,  and the last line defines the operator $\mathcal{O} $.  Using the $r$ dependent variable $\theta_R$ of \cite {fuller} would have eliminated the $\partial/\partial \cos \theta$ term in $ \mathcal{O} $, but made the $\nu-\nu$ interaction very difficult to deal with when the $\nu$-sphere radii depend on flavor. Results the equivalent of (\ref{angles}) are found in the appendix to ref.\cite{sph}.
The $\vec \sigma_\alpha$'s are related to the flavor-density matrices, $\vec  \Phi_{\alpha} ({r},\cos \theta)$, for individual neutrinos by 
\begin{eqnarray}
\vec \sigma_{\alpha}(r,\cos \theta)=\sigma_{0,\alpha} (r,\cos \theta)\,\vec  \Phi_{\alpha} ({r},\cos \theta)\,, 
\end{eqnarray} 
where $\sigma_{0,\alpha} ({ r},\cos \theta)$ is the total neutrino density in the group $\alpha$.
We have $ \mathcal{O}  \vec \sigma_{\alpha} ({ r},\cos \theta)=0$ in the absence of $\nu$ interactions, and  $ \mathcal{O}  \sigma_{0,\alpha} ({ r},\cos \theta)=0$
even in the presence of the neutrino terms, so that (\ref{master0}) becomes,
\begin{eqnarray}
& \mathcal{O} \vec \Phi_{\alpha} ({ r},\cos \theta)   = 2^{1/2}(2\pi)^{-3}G_F 
\vec \Phi_{\alpha} ({ r},\cos \theta)\, \times
\nonumber\\
& \times\sum_\beta \zeta_\beta  \int d\Omega_{\bf \hat n'} \sigma_{0,\beta} ({r},\cos \theta') \vec  \Phi_{\beta}  ({\bf r},{\bf \hat n'})
(1-{\bf \hat n} \cdot {\bf \hat n'} ) \,.
\label{master2}
\end{eqnarray}

In the flavor diagonal initial state we have $ \Phi^{(z)}_\alpha =1$, $ \Phi^{(x),(y)}_\alpha =0$, for all $\alpha$. (Here and in what follows the superscripts $x,y,z$ stand for directions $1,2,3$ in flavor space.)
The solutions to $ \mathcal{O}  \sigma_{0,\alpha} ({\bf r},\hat n)=0$ are of the 
form $\sigma_{0,\alpha} =f_\alpha (r \sin \theta)$, where the $f_\alpha$'s are any functions. 
The ``neutrino bulb" model assumes $\sigma_{0,\alpha}=n_\alpha \Theta (r_0-r \sin {\theta})$, independent of $\alpha$, where $\Theta$ is the Heaviside function and $n_\alpha$ are the number densities of the species $\alpha$ on a common surface. We now
relax this assumption by taking different effective neutrino-surfaces for the different groups of neutrinos,
 \begin{eqnarray}
\sigma_{0,\alpha}=n_\alpha \Theta (r_{\alpha}-r \sin \theta)\,,
\nonumber\\
\label{B}
\end{eqnarray} 
where $r_\alpha=\{r_{\nu_e},r_{\bar \nu_e},r_{\nu_x},r_{\bar \nu_x}\}$, with $r_{\nu_e}$ being the largest. Next we re-express the right hand side of (\ref{master2}) in a series of steps:\newline
a). We replace $\cos \theta$ by a new variable $s$ with the relation chosen differently for each of the four functions $\Phi_{\alpha}(r,\cos\theta)$, where
$\cos \theta =1-s  \, h_\alpha (r)$, and 
\begin{eqnarray}
h_\alpha(r)=1-[1-(r_{\alpha}/ r)^2]^{1/2}\,.
\end{eqnarray}
b) Step a) above entails the replacement in (\ref{master2}),
\begin{eqnarray}
&\int d \Omega_{\hat n'}\sigma_{0,\beta} ({r},\cos \theta') \vec  \Phi_{\beta}(r,\cos\theta')( 1-\hat n\cdot \hat n') \rightarrow 
\nonumber\\
 & 2\pi n_\beta \int_0^1 ds'  \vec   \Phi_{\beta} (r,s')h_\beta ( h_\alpha s+h_\beta s'- h_\alpha h_\beta s s')
\end{eqnarray}
c). In the new variables we have, 
\begin{eqnarray}
&\mathcal{O}=
[1-s\, h_\alpha(r )]{\partial \over\partial\,r} + \Bigr [{h_\alpha(r) s^2-2 s \over r}+
\nonumber\\
&{(s^2 h_\alpha(r)-s) h_\alpha'(r)\over h_\alpha(r)}\Bigr ]{\partial \over\partial s}\,,
\label{O2}
\end{eqnarray}
 and the evolution equation is,
\begin{eqnarray}
&\mathcal{O}\vec\Phi_\alpha (r,s)=
\nonumber\\
& 2^{-3/2}\pi^{-2} G_F   \vec\Phi_\alpha (r,s,)\times \vec u_\alpha (r,s )  \,,
\label{correct3}
\end{eqnarray}
where
\begin{eqnarray}
&\vec u_\alpha (r,s )=\sum_{\beta}  n_{\beta}\zeta_{\beta} h_\beta
  \int_{0}^1 ds' [s h_{\alpha}+s' h_{\beta} -
  \nonumber\\
  &s s' \,h_{\alpha}h_{\beta}] \vec \Phi_{\beta}(r,s')\,.
\nonumber\\
\label{prod}
\end{eqnarray}
We shall examine instabilities by making the perturbation $\Phi \rightarrow \Phi+\chi$ , then linearizing the RHS of (\ref{correct3}) in 
$\chi$. In all calculations we shall discretize in the variables $s,s'$ with N equal steps in the interval $(0,1)$.

This calculation will be local, in the sense that the instability criteria will involve the background flavor-density matrix
only at a single point, and for that we use the results of free-streaming from the several separate $\nu$-surfaces. 
When we then look to plot the growth of instabilities, we concern ourselves with a volume of surrounding space 
small enough such that in the absence of instabilities there would have been no appreciable variations of $\Phi$.
From (\ref{correct3}) we obtain,
\begin{eqnarray}
&\mathcal{O} \vec \chi_{\alpha,i}={2 ^{-3/2}G_F\over \pi^2  N} \Bigr [  \vec \chi_{\alpha,i}\times \sum_{\beta=1}^4 \sum_{j=1}^N 
\vec \Phi_{\,\beta,j} w_{\alpha,i;\beta,j} + 
\nonumber\\
& \vec \Phi_{\alpha,i} \times  \sum_{\beta=1}^4 \sum_{j=1}^N \vec  \chi_{\,\beta,j} w_{\alpha,i;\beta,j} \Bigr],
\label{perto}
\end{eqnarray}
where $w_{\alpha,i;\beta,j}$, with discretized angular indices $(i,j)$, is given by
\begin{eqnarray}
w_{\alpha,i;\beta,j}=h_{\beta}n_{\beta} \zeta_{\beta} [i \,h_\alpha N^{-1} +j \, h_\beta  N^{-1}+i  \,j\,h_\alpha  \, h_\beta N^{-2}].
\nonumber\\
\,
\end{eqnarray}
In all that follows, we take $\Phi^{(z)}_{\alpha, i}=1$,  $\Phi^{(x),(y)}_{\alpha, i}=0$ in (\ref {perto}) for all $\{\alpha, i\}$, since the initial state is to have seen no neutrino oscillation, giving,
\begin{eqnarray}
\mathcal{O}  \, \chi_{\alpha,i}^{(x)}= \sum_{\beta,j} M_{\alpha,i;\beta,j}\,  \chi_{\beta,j}^{(y)}\,,
\nonumber\\
\, \mathcal{O} \, \chi_{\alpha,i}^{(y)}=- \sum_{\beta,j} M_{\alpha,i;\beta,j}\, \chi_{\beta,j}^{(x)}\,,
\label{perts}
\end{eqnarray}
where,
\begin{eqnarray}
&M_{\alpha,i;\beta,j}=2^{-3/2} \pi^{-2} G_F N^{-1}\times
 \nonumber\\
& \Bigr[ \delta_{\alpha,\beta}\, \delta_{i,j} \sum_{\beta',j'} w_{\alpha,i; \beta',j'} - w_{\alpha,i;\beta,j}   \Bigr ].
  \label{mat}
 \end{eqnarray}
The indices$\{\alpha,i\}$ and $\{\beta,j \}$ henceforth will be consolidated into $a$ and $b$,
respectively, in a $4N$ dimensional
space. The eigenvalues
of $M_{a,b}$, as a matrix in this space, then indicate instability if any have an \underline{imaginary} part. (It is the $x\leftrightarrow -y$ symmetry in (\ref{perts}) that converts an imaginary part here effectively into a real part for an eigenvalue of the system as a whole.) The coupled equations (\ref{perts}) 
would then have exponentially increasing solutions, as long as an initial value has any component in the direction of the eigenvector.
The first term on the RHS of {(\ref{mat}) is diagonal and real but the second term is non-hermitian; nonetheless for a wide range of parameters $r_{\alpha}$ and $n_\alpha$ that we have considered we find stability. We believe this stability with respect to spherically symmetric perturbations to be a general result.

To address azimuthal instabilities at the initial surface $r=r_{\nu_e}$ we take the unperturbed $\Phi_{\alpha,i}$ as given by our previous expressions, and therefore with angular distributions
that are functions of $h_\alpha( r_{\nu_e})$ as above. For the linear azimuthal perturbation we take the form $ \Delta \Phi^{ \{ x,y \} }= \chi^{ \{ x,y \}} [\cos \theta)] \, \cos\phi$, where the azimuthal angle $\phi$ is defined locally with respect to a tangent to the surface. 
Now in the analogue to the final term in (\ref{perto}), performing the $\phi'$ integral over $\hat n\cdot \hat n' $, we obtain,
\begin{eqnarray}
&\mathcal{O} \chi_{\alpha,i}^{(x)}=2^{-3/2}\pi^{-2} N^{-1}G_F \Bigr [   \chi_{\alpha,i} ^{(y)} \sum_{\beta,j} \Phi_{\,\beta,j} ^{(z)}w_{\alpha,i;\beta,j} +
\nonumber\\
&{1\over 2} \Phi_{\alpha,i}^{(z)}  \sum_{\beta,j}\chi_{\,\beta,j}^{(y)}\,\zeta_\beta \,n_\beta\, h_\beta\,
   v_{\alpha,i} \,v_{\beta,j} \Bigr]
\label{perto2}
\end{eqnarray}

where the discretized $\sin \theta$ factors from $\hat n\cdot \hat n' $ have the form,
$v_{\alpha,i}(r)= [1-(1-i h_\alpha (r)/N)^2]^{1/2}$.
Following the previous development we then recapture (\ref{perts}), but with $M$ replaced by,
\begin{eqnarray}
 \tilde M_{\alpha,i;\beta,j}
 =2^{-3/2} \pi^{-2}G_F N^{-1}[ w^{(1)}_{\alpha,i;\beta,j} +w^{(2)}_{\alpha,i;\beta,j}   ],
 \label{mat2}
 \end{eqnarray}
 where
 \begin{eqnarray}
 w^{(1)}_{\alpha,i;\beta,j}=\delta_{\alpha,\beta}\, \delta_{i,j} \sum_{\beta',j'}   w_{\alpha,i; \beta',j'}\,,
 \end{eqnarray}
 and
 \begin{eqnarray}
w^{(2)}_{\alpha,i;\beta,j}=2^{-1} \, \zeta_{\beta}\,  n_\beta\,h_{\beta}\, v_{\alpha,i}\,  v_{ \beta,j}\,.
\end{eqnarray}

We shall look for instabilities at $r=r_{\nu_e}$, the outermost neutrino-surface. Their presence now depends on the values of four parameter ratios, 
$r_{\bar\nu_e} /r_{\nu_e}$, $r_{\nu_x} /r_{ \nu_e}$, $n_{\bar \nu_e}/n_{ \nu_e}$,  $n_{\nu_x}/n_{ \nu_e}$. Explosion calculations give rather definite values for the number density ratios, which can be calculated from the ratios of luminosity to average energy. The raw ratios use the luminosities given by the simulations at a common radius outside all neutrino spheres, but still in the region in which neutrino oscillations matter very little. But the $n_\alpha$ are the densities at the respective $\nu$-surfaces; and we must therefore apply a correction factor $[r_{ \nu_e}/r_{\alpha}]^2$ to the raw ratios for $\alpha=\bar \nu_e,\nu_x, \bar \nu_x$.

Here we shall use the values given in fig. 7 of ref. \cite{janka-rev} in the range of times, .3 sec.-1 sec., post-bounce.
We find  $n_{\bar \nu_e}/n_{\nu_e}\approx .77 [r_{ \nu_e}/r_{ \bar \nu_e}]^2$, $n_{\nu_x}/n_{\nu_e}\approx .62[ r_{ \nu_e}/r_{\nu_x}]^2$.  We have not found as 
authoritative predictions for neutrino-surface radii; a neutrino-surface is an idealization that is not contained in the data. But we believe that, given the idealization, the opacities and the matter density profile dictate that $r_{\nu_x}<r_{\bar \nu_e}<r_{\nu_e}$. Within this range and for the above values of $n_\alpha$ ratios, we find instability
in a region approximately described as,
\begin{eqnarray}
r_{\bar \nu_e}/r_{  \nu_e}>.44+.55\, r_{ \nu_x}/r_{ \nu_e}\,.
\label{region}
\end{eqnarray}
 Thus if the radius for $\nu_x$ is 
80\% of the radius for $ \nu_{e}$ then only the region
$r_{\nu_{e}}>r_{\bar\nu_{e}}>.88\, r_{\nu_{e}}$ is unstable.

The instability is manifested by imaginary parts in a single physical pair of eigenvalues,
 $\lambda, \lambda^*$, of $\tilde M$ converging rapidly to a limit as we increase the subdivision number N (say, from 16 to 128), while spurious imaginary parts in other eigenvalues as a result of finite subdivision 
(noted in \cite{raf8}) go to zero as $N^{-1}$. As an example, when we choose $n_{\bar \nu_e}=10^{32}{\rm{cm}^{-3}}$ at  $r=15 \,{\rm km}$, and choose the ratios $r_{\bar\nu_{e}}=.93\, r_{\nu_e}$, $r_{\bar\nu_{x}}=.8\, r_{\nu_e}$ 
we find ${\rm Im}[\lambda]=\pm .008  G_F n_{\bar \nu_e}\approx .32 {\rm}{\rm \,(meter})^{-1}$.

The critical region given by (\ref{region}) appears to be likely to be realized at some times or places during the event.
As to ``places", we suggest a broader context for our considerations. The interior simulations suggest an environment that is locally very irregular in its parameters of density and temperature. This will translate into very irregular neutrino angular distributions as well. 
If, as in simulations like those of  \cite{fuller}-\cite{fuller9}, the important flavor evolution takes place over a region in $r$ many times the size of the $\nu$-surface, these irregularities may get averaged out, since at most points on the way out we see solid angle of nearly $2 \pi$ of the $\nu$-surface.
But if the scale of $\nu$ flavor exchange is tens of meters instead, then we see only a small fraction of the complete $\nu$-surface as we sit only a little ways above it, and the averaging argument is less sustainable.
We regard our present work as a step of interpolation between the methods and results for the usual smoothed-out model and those of a future statistical model. 

For the remainder of the paper we shift to taking the system 
at time $t=0$ to have been established in a state of steady symmetric flow, and we then set $\mathcal{O}=d/dt$, turn on the perturbation and follow the time development at a fixed point, seeking the results for some period of time short compared to $r_0$ but long compared to the exponential 
rise times.  Following that path we now examine the
back-reaction on $\Phi^{(z)}$ of the growing transverse modes, $\chi^{(x),(y)}$ . We use the complex representation,
$\chi_a=\chi_a^{(x)}+i\chi_a^{(y)}$ in what follows.
Taking the normalized eigenvector of $\tilde M$ for the growing mode to be $\xi_a$, with eigenvalue $\lambda$, we assume $ \chi_{a}(t)\approx  g(t) \xi_a$, for all $a\equiv (\alpha,i)$, that is, we take the growing mode to dominate the $(x,y)$ components. The evolution equation (\ref{master2}) couples all of the $\Phi_a^{(z)}$ modes to the perturbations, $\chi$.  Defining $z(t)=(4N)^{-1} \sum_a \Phi_a^{(z)}$ (i. e. summed over all angles and all flavors), we find,
 
\begin{eqnarray}
&   \dot z(t)=2^{-1/2} i G_F n_\nu |g(t)|^2 \sum_{a ,b} \xi_{a}  \Bigr ([w^{(2)}]^\dagger_{a,b}-w^{(2)}_{a,b}\Bigr )\xi_b
   \nonumber\\
&   = |g(t)|^2 {\rm Im}[\lambda]\,, 
 \label{eveq}
   \end{eqnarray} 
where to get the second form we first replaced 
$w^{(2)}$ by $w^{(1)}+w^{(2)}$, since $w^{(1)}$ is Hermitian, then used the fact that $\xi_a$ is an eigenvector of $w^{(1)}+w^{(2)}$, with
 $\lambda$ the  growing mode eigenvalue of the matrix $\tilde M$ of (\ref{mat2}). This is supplemented with,
 \begin{eqnarray}
 \dot g(t)=i  \lambda z(t) g(t) \label{g(t)}
 \label{gdot}
 \end{eqnarray}
which comes directly from (\ref{perts}). 
Using (\ref{gdot}) we obtain the first equality in
\begin{eqnarray}
{d \over dt}[|g(t)|^2]=-2 {\rm Im}[\lambda] z(t) |g(t)|^2=-2 z(t)  \dot z(t) 
\end{eqnarray}
the second following from (\ref{eveq}), and giving
the solution $|g(t)|^2=-z(t)^2+C$. We take $z(0)=1$, so that $C=|g(0)|^2 +1$. 
Now differentiating (20) and using the above we obtain,
\begin{eqnarray}
 \ddot z(t)= 2 [{\rm Im} \lambda]^2 [z(t)^2- (1+ |g(0)|^2 )] z(t)\,.
 \label{zap}
\end{eqnarray}
 The parameter $|g(0)|^2$ is the square of the initial $\nu$ mixing amplitudes, from whatever source below in the star, in particular the one discussed in 4) in the introduction.
In fig. 1 we plot the results for values  $|g(0)|^2=10^{-4}, \,10^{-6}, \,10^{-8} $, using the eigenvalue calculated previously to express the scale in $\rm meters/c$.

\begin{figure}[h] 
 \centering
\includegraphics[width=2.5 in]{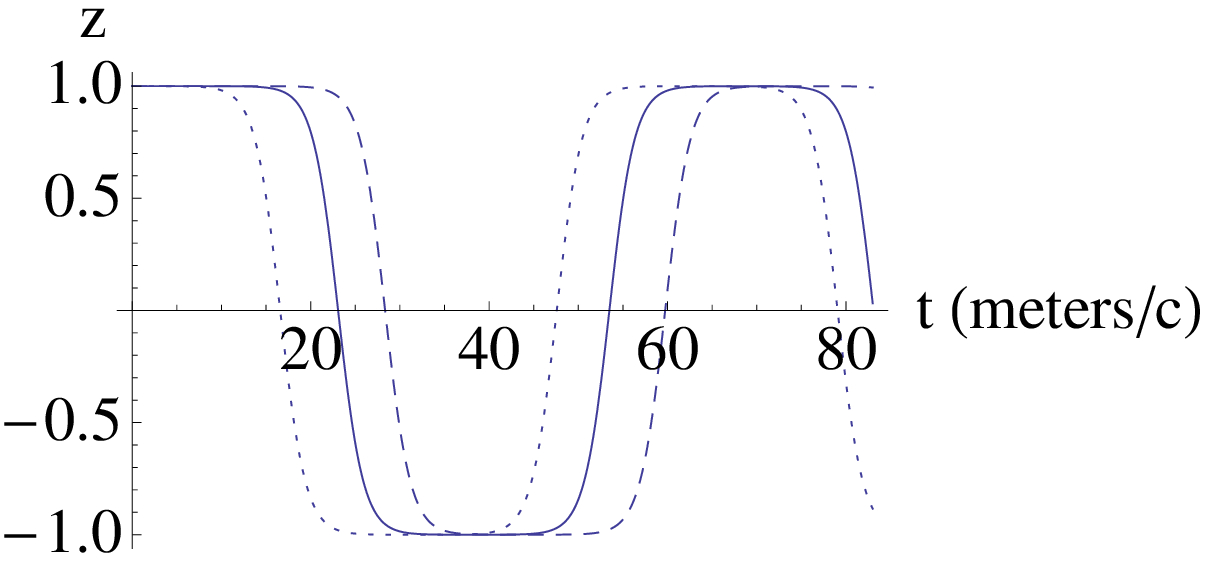}
 \caption{ \small }
Flavor turnovers as measured by $z(t)$, plotted against time.  The dotted, solid, and dashed curves are for the cases  
 $|g(0)|^2=10^{-4}, \,10^{-6}, \,10^{-8} $, respectively.
  \label{fig. 1}
\end{figure}
Recall that in our definition, $\Phi^{(z)}_{\alpha,i} $ tracks the evolution in the stream that has original flavor $\alpha$. 
 A transition in which $(4N)^{-1}\sum_a\Phi_a^{(z)}$ goes from $1$ to $-1$ as in fig.1 means complete $e\leftrightarrow x$ exchange for each beam.  
The curve during the short transition is almost a perfect ``$\tanh$" kink as long as the initial values for $|g(0)|^2$ are sufficiently small that the length of the initial plateau is long compared with the transition time, as in the plotted cases. Note that the plots begin at the middle of a plateau and the turn-over time is proportional to $|(\log |g(0)|)|^{-1}$. For each reduction of a factor of 10 in the size of the seed, $g(0)$, the turn-over is 2.5 meters farther away from the initial surface. 
This observation is important in dealing with the question of the matter effects, which in other calculations damp the growth of oscillations in the regions of higher electron density. In our calculation they barely matter; a tiny push, 
$|g(0)|^2$, gets the unstable system moving, but damping this push by a factor of 100 hardly changes the results.
The non-linear interaction exchanges the directions and energies of e and x neutrinos without changing their numbers, and likewise for the $\bar \nu$'s. The total interaction energy with electrons in the medium therefore does not change in this transformation since it depends just on densities and not on spectra.

This is quite different from the situations in other literature in which one starts on a single surface in a flavor-diagonal state with a species independent angular distribution and finds instabilities only after a substantial amount of mixing from the oscillation terms. At the shorter distances (say 15 km) with high electron densities this mixing is strongly suppressed by the interaction with electrons. This is why in typical calculations one gets sudden turnovers only at distances of order 100 km.

Fig. 1 shows total  $\nu_e \leftrightarrow \nu_x$ interchange over a very short time from the beginning of the calculation (for a case where $\nu_e=\nu_x$). Since the temperature ratios at the $\nu$-surface are expected to be of the order $T_{\nu_x}/T_{\nu_e}\approx 2$ this would represent a 100\% heating for the $\nu_e$, which in turn would much more efficient at heating the matter in this region. This sudden heating in the neutrino-surface region would be expected to be very important to the explosion calculation.
We can only speculate about what might happen in a more realistic case in which the gross governing feature remains the differing neutrino-surface radii among the species, but the seeds (still on a nearly planar local initial surface) are more irregular on that surface, e. g. in $\phi$ dependence. As we move outward the different domains originating with different areas on the original surface will begin to merge.
All of this appears most likely to lead to a rather quick averaging in which we settle at $z\approx 0$ on average in fig. 1. Then instead of a 100 \% gain in temperature of $\nu_e$ 's we would have in effect a 50\% increase, still very potent in terms of heating the surrounding matter. 

We believe that this last possibility is sufficiently likely to justify a complete explosion calculation in which the flavor dependence of neutrino distributions is homogenized computationally, with respect to flavor-energy distributions,  at frequent intervals, as the calculation proceeds outward through the neutrino-sphere region. The object would be to test qualitatively whether or not an extension of the considerations of the present paper can have an important effect on explosion dynamics

It is interesting that the structure of the equations that enter the above calculations is very similar to those for non-linearly interacting photons, where now flavor is to be replaced by photon polarization in real space. This is another reason to study general behaviors and computational algorithmic development. Possible applications range from laboratory studies in media \cite{rfs2} to the polarization signal in the prompt emission from gamma ray bursts. In the latter case the vacuum $\gamma-\gamma$ interaction remains effectively strong out to regions of small optical depth, so that the data could be useful in understanding the source.

I am much indebted to Georg Raffelt for making a key observation at the very beginning of this work.

 \end{document}